\documentclass[12pt]{spieman}  
\usepackage{amsmath,amsfonts,amssymb}
\usepackage{graphicx}
\usepackage{setspace}
\usepackage{lineno}
\usepackage{tocloft}

\title{Measurement of the Temperature Dependence of the Refractive Index of CdZnTe}

\author[a,b,*]{Umi Enokidani}
\author[a,b]{Hideo Matsuhara}
\author[b,c,h]{Takao Nakagawa}
\author[b]{Shunsuke Baba}
\author[d]{Yasuhiro Hirahara}
\author[e]{Ryoichi Koga}
\author[d]{Yuan LI}
\author[d]{Biao Zhao}
\author[a,b]{Daiki Takama}
\author[d,f]{Hiroshi Sasago}
\author[g]{Takehiko Wada}

\affil[a]{The Graduate University for Advanced Studies, Shonan Village, Hayama, Kanagawa, Japan}
\affil[b]{Institute of Space and Astronautical Science, Japan Aerospace Exploration Agency, Chuo-ku, Sagamihara, Kanagawa, Japan}
\affil[c]{Department of Physics, Graduate School of Science,The University of Tokyo, Hongo, Bunkyo-ku, Tokyo, Japan}
\affil[d]{Nagoya University, Furocho, Chikusa-ku, Nagoya, Aichi, Japan}
\affil[e]{Nagoya City University, Mizuho-cho, Mizuho-ku, Nagoya, Aichi, Japan}
\affil[f]{Sasago Co., Ltd., Nagoya, Aichi, Japan}
\affil[g]{National Astronomical Observatory of Japan, Osawa, Mitaka, Tokyo, Japan}
\affil[h]{Tokyo City University, Tokyo, Japan}

\cftpagenumbersoff{figure}
\cftpagenumbersoff{table} 
\begin{document} 
\maketitle

\begin{abstract}

We have been developing a CdZnTe immersion grating for a compact high-dispersion mid-infrared spectrometer (wavelength range 10--18 $\mu$m, spectral resolution $R = \lambda/\Delta \lambda > 25,000$, operating temperature $T < 20$ K). Using an immersion grating, the spectrometer size can be reduced to $1/n$ ($n$: refractive index) compared to conventional diffraction gratings. CdZnTe is promising as a material for immersion gratings for the wavelength range. However, the refractive index $n$ of CdZnTe has not been measured at $T < 20$ K.

We have been developing a system to precisely measure $n$ at cryogenic temperatures ($T \sim 10$ K) in the mid-infrared wavelength range. As the first result, this paper reports the temperature dependence of $n$ of CdZnTe at the wavelength of 10.68 $\mu$m. This system employs the minimum deviation method. 
The refractive index $n$ of CdZnTe is measured at temperatures of \( T = 12.57, 22.47, 50.59, 70.57, \text{ and } 298 \, \text{K} \). We find that $n$ of CdZnTe at $\lambda =$ 10.68 $\mu$m is $2.6371 \pm 0.0022$ at $12.57 \pm 0.14$ K, and the average temperature dependence of $n$ between 12.57 $\pm$ 0.14 K and 70.57 $\pm$ 0.23 K is $\Delta n/\Delta T = (5.8 \pm 0.3) \times 10^{-5}$ K$^{-1}$. 

\end{abstract}


{\noindent \footnotesize\textbf{*}Umi Enokidani,  \linkable{enokidani.umi@jaxa.jp} }

\begin{spacing}{2}   

\section{Introduction}
\label{sect:intro}  

\begin{figure}[H]
\begin{center}
\begin{tabular}{c}
\includegraphics[height=5.5cm]{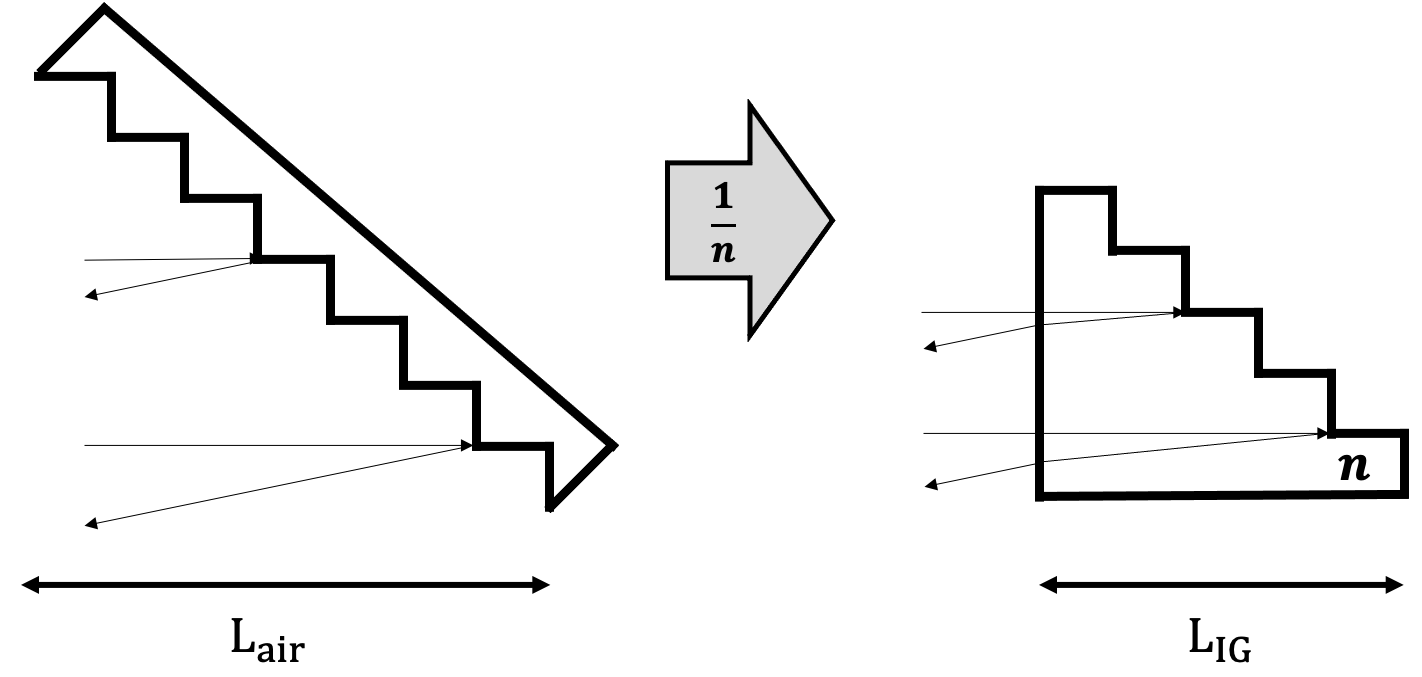}
\end{tabular}
\end{center}
\caption 
{ \label{fig:immersion grating}
Conceptual view of the immersion grating. Conventional diffraction grating (left) and immersion grating (right)} 
\end{figure}

The mid-infrared region of 10--18 $\mu$m, which is difficult to observe from the ground, contains many interesting molecular transitions, such as those of water, carbon dioxide, and ammonia. 
The H$_2$O snowline is the boundary where water changes between gas and solid. The H$_2$O snowline is a critical factor separating the formation of rocky and gaseous planets. By observing the gas distribution, the location of the snowline can be determined. To measure the gas distribution, spectroscopy is used instead of spatially resolved observations. By assuming that the gas moves in Keplerian motion, its spatial distribution can be obtained. The edge of this distribution corresponds to the snowline.
For example, previous studies suggested that the H$_2$O snowline in a protoplanetary disk can be identified by observing the Keplerian motion of optically thin mid-infrared H$_2$O gas emission lines, such as the 17.754~$\mu$m line \cite{Notsu_2016, Notsu_2017}.
To realize this observation, it is necessary to achieve high optical efficiency measurements with $R>$ 25,000 for the specific H$_2$O line at 17.754 $\mu$m \cite{Nakagawa_2024}. However, even NASA's flagship space telescope, JWST (James Webb Space Telescope), with its Mid-Infrared Instrument (MIRI), which covers 17.754 $\mu$m line, has a spectral resolution limited to 3,250 \cite{Argyriou_2023}. 
The spectral resolution $R$ is proportional to the optical path difference, and thus high R conventionally requires  large diffraction gratings in high-resolution spectrometers. In optical systems using traditional gratings, it is difficult to achieve \( R > 25,000 \) due to limited volume and cooling resources when these systems are mounted on space telescopes. \cite{Kobayashi_2008, Sarugaku_2012}
For these reasons, we have been developing a reflective-type dispersive element, the immersion grating, for the mid-infrared range. 
For the next-generation infrared space telescope GREX-PLUS (Galaxy Reionization EXplorer and PLanetary Universe Spectrometer)\cite{Inoue_2024} and for its compact high-dispersion mid-infrared spectrometer, HRS (High Resolution Spectrometer)\cite{Nakagawa_2024}, we have been developing a CdZnTe immersion grating.
GREX-PLUS/HRS operates in the wavelength range of 10--18 $\mu$m, with $R > 25,000$ and an operating temperature of $T < 20$ K.
An immersion grating is an optical element that can reduce the size of the spectrometer to $1/n$ ($1/n^3$ in volume, $n$: refractive index) compared to conventional diffraction gratings (see Fig \ref{fig:immersion grating}). This is because passing light through a high-refractive-index medium allows for an effectively longer optical path length $L_{\text{IG}}$ compared to $L_{\text{air}}$ in air by a factor of $n$.

So far, there have been demonstrations of immersion gratings for the near-infrared, such as the Si immersion grating used in IGRINS (Immersion GRating INfrared Spectrometer) \cite{Marsh_2007, Gully-Santiago_2014} and the Ge immersion grating used in GIGMICS (Germanium Immersion Grating Mid-Infrared Cryogenic Spectrograph) \cite{Hirahara_2010}. However, no immersion grating has yet been developed for the mid-infrared region of 10--18 $\mu$m.

The material for the immersion grating should have a high $n$ and transmittance \( \tau_{\text{sum}} \) in the observation wavelength range and at the operating temperature of $T< 20$ K. It also requires enough strength for fine processing. From these perspectives, CdZnTe is selected as the material for the immersion grating \cite{Sarugaku_2017}. However, no literature values are available for $n$ of CdZnTe in the mid-infrared range, and the value currently used for planning GREX-PLUS/HRS was estimated based on an analogy with CdTe, which belongs to the same cubic system and has a similar elemental composition \cite{Hlidek_2001}. Also, they are supposed to not exhibit significant birefringence. The accurate $n$ for CdZnTe is crucial for determining the wavelength dispersion capability of the immersion grating. It is also essential for the detailed design of the spectrometer’s optical system to achieve high efficiency at a target wavelength. This accuracy is required as follows. First, the maximum resolution $R_{\text{max}}$ of the immersion grating at wavelength $\lambda$ under the Littrow condition is given as \cite{Ikeda_2015}

\begin{equation}
R_{\text{max}} = \frac{2n\phi \tan \theta_{\text{bla}}}{\lambda},
\label{eq:resolution}
\end{equation}
where $\phi$ is the diameter of the collimated beam, and $\theta_{\text{bla}}$ is the blaze angle. The blaze wavelength $\lambda_{\text{bla}}$ is given by

\begin{equation}
\lambda_{\text{bla}} = \frac{2n\sigma \sin \theta_{\text{bla}}}{m} 
\label{eq:blaze wavelengh}
\end{equation}
and the free spectral range $\Delta \lambda$ is given by
\begin{equation}
\Delta \lambda \sim \frac{\lambda_{\text{bla}}}{m} \quad (m \gg 1),
\label{eq:free spectral range}
\end{equation}
where $\sigma$ is the grating pitch, and $m$ is the diffraction order, which is considered to be $m \approx 100$ for GREX-PLUS / HRS due to the constraints of the detector pixel format. The blaze wavelength $\lambda_{\text{bla}}$ corresponds to the peak of diffraction efficiency within $\Delta \lambda$. 
The decrease in diffraction efficiency must be limited to 10\% so that the 17.754~$\mu$m H$2$O gas emission line remains within 0.18$\Delta \lambda$ around $\lambda_{\text{bla}}$.
To achieve this limit with 3$\sigma$ confidence, it is necessary to measure $n$ with uncertainties less than $\Delta n < 10^{-3}$.



In addition, accurate values of $n$ are required to precisely determine the absorption coefficient. To determine the absorption coefficient, the transmittance of two samples with different thicknesses is measured. 
The measurements must be corrected for the Fresnel reflection loss caused at the front and back surfaces of each sample. The Fresnel reflectance is given as
\begin{equation}
R_{\textbf{fre}} = \left( \frac{n - 1}{n + 1} \right)^2.
\label{eq:Frenel}
\end{equation}
Therefore, accurate value of $n$ is necessary to robustly correct the Fresnel loss component. Li $et$ $al.$ (2024) \cite{Li_2024} derived the absorption coefficient, considering $R_{\text{fre}}$, estimated from a provisional value of $n$ obtained from our study.

\section{Experimental Setup}
\label{sect:Experimental Setup}  
In this study, we adopt the minimum deviation method \cite{born_wolf_optics} to measure $n$. This method allows accurate measurements and is regarded as a standard measurement technique. This method can directly measure the absolute value of $n$ and has been employed for low-temperature refractive index measurements e.g., in CHARMS \cite{Leviton_2022} at NASA/Goddard Space Flight Center and the measurements by Yamamuro et al. et al. 2006.\cite{Yamamuro_2006}

\subsection{Measurement principle}
\begin{figure}[H]
\begin{center}
\begin{tabular}{c}
\includegraphics[height=8.0cm]{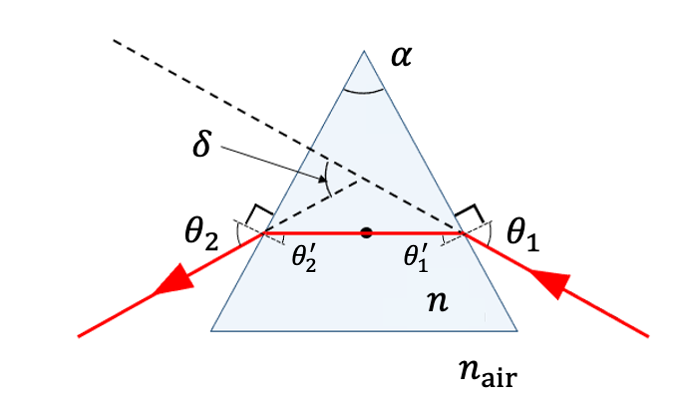}
\end{tabular}
\end{center}
\caption 
{ \label{fig:prism_refraction}
Diagram illustrating the refraction of a monochromatic collimated beam passing through a prism with an apex angle $\alpha$. 
The refractive index of the prism's material is denoted as $n$, and the refractive index of air is $n_{\text{air}}$. The incident and exit angles are \( \theta_1 \) and \( \theta_2 \), the corresponding refraction angles are \( \theta'_1 \) and \( \theta'_2 \), and the deviation angle is $\delta$.}
\end{figure} 

Fig \ref{fig:prism_refraction} shows monochromatic collimated light entering a prism with $\alpha$ and $n$ at $\theta_1$. After refraction, the light exits at an exit angle $\theta_2$. The angle between $\theta_1$ and the exit angle $\theta_2$ is $\delta$. When $\theta_1$ is equal to the exit angle $\theta_2$, the deviation angle $\delta$ reaches its minimum value, which is referred to as the minimum deviation angle $\delta_{\text{min}}$.The refractive index $n$ can be determined using $\delta_{\text{min}}$ and $\alpha$ of the prism through the following 

\begin{equation}
n = \frac{\sin \left( \frac{\delta_{\text{min}} + \alpha}{2} \right)}{\sin \left( \frac{\alpha}{2} \right)}.
\label{eq:minimum_deviation_method}
\end{equation}
Therefore, we first optically measure $\alpha$ of the prism with precision exceeding the mechanical processing accuracy. Then, we find $\delta_{\text{min}}$ by varying $\theta_1$.

\subsection{Optical system}
\label{Optical system}

\begin{figure}[H]
\begin{center}
\begin{tabular}{c}
\includegraphics[height=8.0cm]{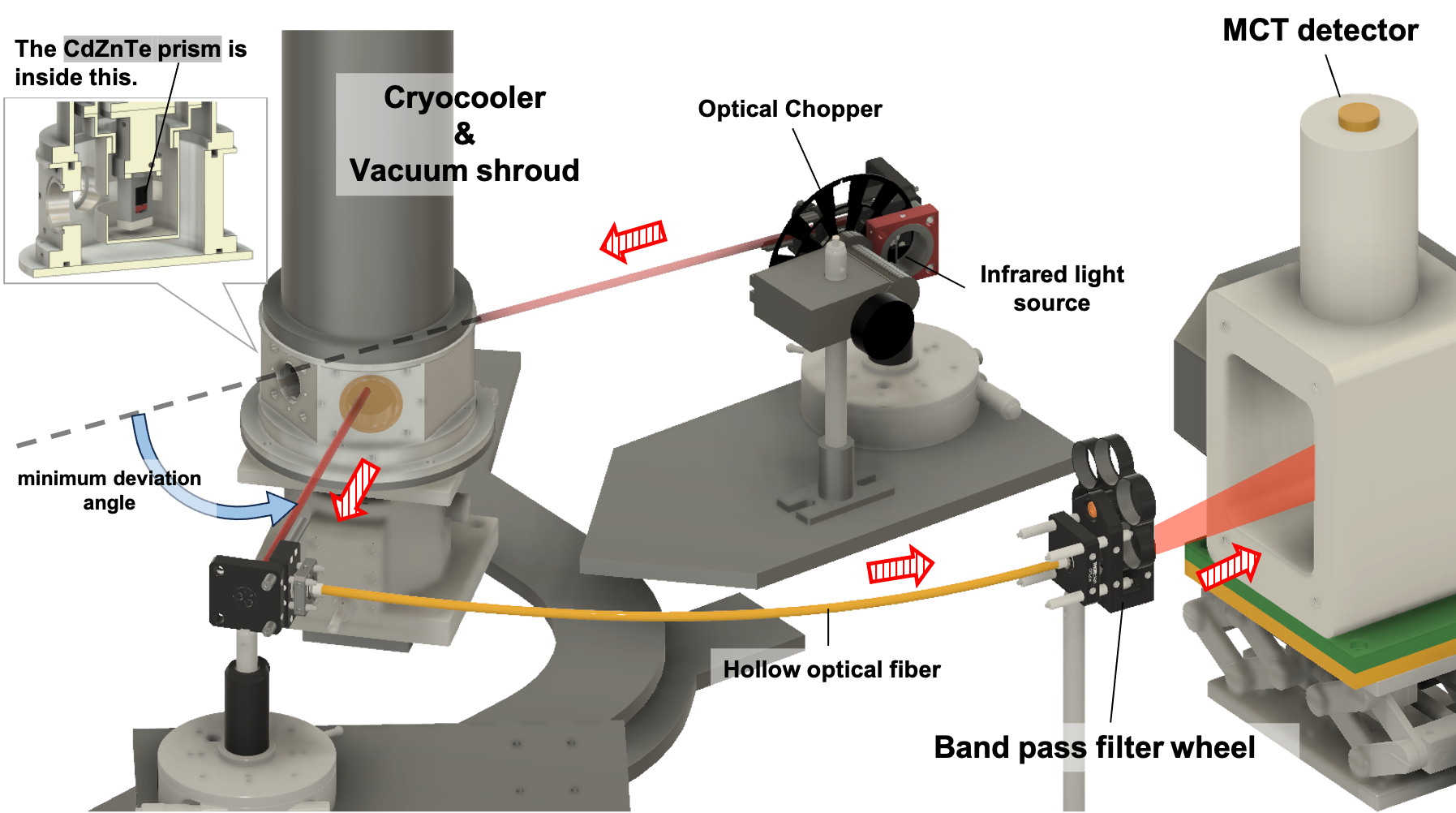}
\end{tabular}
\end{center}
\caption 
{ \label{fig:3D_configuration_system}
3D configuration of the cryogenic infrared refractive index mesurement system.
 The red arrow indicates the direction of the measurement light. How this system is used in the apex angle measurement and the deviation angle measurement is shown in Figs. \ref{fig:autocollimation_setup} and \ref{fig:deviation_setup}, respectively.} 
\end{figure} 

We modified a gonio-type refractometer \cite{Enokidani_2024}, the GMR-1, an older model from Shimadzu Device Corporation's GMR series produced by Kalnew-Iida Works (Fig \ref{fig:3D_configuration_system}). At the center of the refractometer, is a rotating sample stage where a Cd$_{0.96}$Zn$_{0.04}$Te prism is placed. There are two platforms for placing the optical system: one is a fixed stage, and the other is a movable stage. The movable stage can rotate coaxially with the rotating sample stage. It is equipped with a precise angle encoder (angular resolution of $10^{-4}$ degree), which can measure the rotation angle of the movable stage. Moreover, it allows the rotation of the movable stage to be synchronized with that of the rotating sample stage. In this case, when the movable stage rotates by $\theta$, the rotating sample stage rotates by $\theta/2$. This function enables precise rotation of the rotating sample stage.

On the fixed stage, a filament light source (the black body at approximately 1200 K), a collimator, and an orifice that generates collimated light are placed. Additionally, the fixed stage is equipped with a focusing system that guides light into a fiber, which transmits the light to a detector. There are two types of fibers and detectors: one for visible light and one for infrared light. In front of the infrared detector (MCT), a band-pass filter (BPF) is placed to select the measurement wavelength. A vacuum shroud and a mechanical refrigerator are added to the sample stage (details are explained in the next section). This measurement system allows $n$ to be measured at a wavelength of 10.68 $\mu$m at temperatures between 4 to 300 K. This system also allows similar measurement for a visible wavelength (635 nm) with the detector and filament light source replaced by a photodiode (PD) and a laser, respectively.

\subsection{Cooling system}
\begin{figure}[H]
\begin{center}
\begin{tabular}{c}
\includegraphics[height=6.0cm]{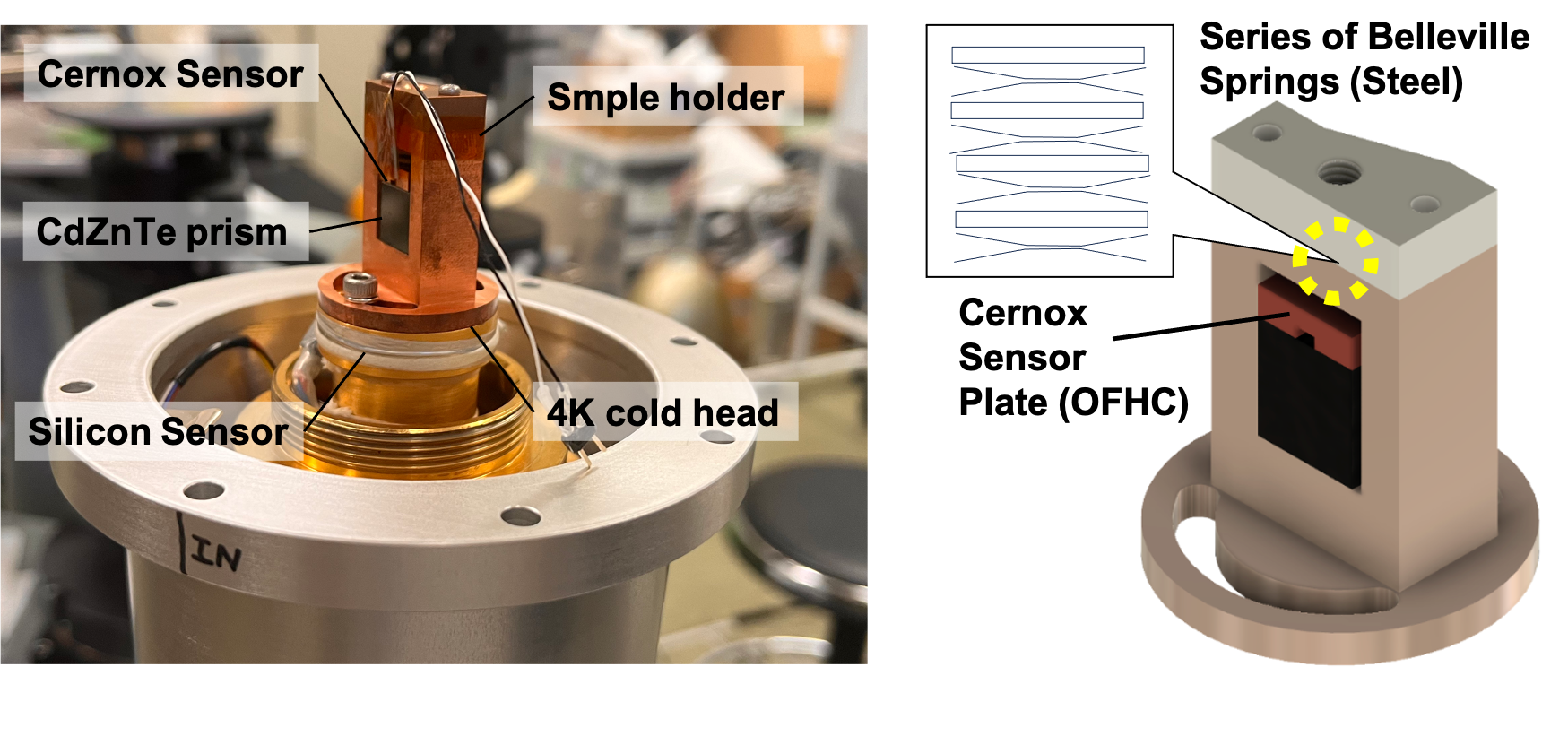}
\end{tabular}
\end{center}
\caption 
{ \label{fig:sample_holder}
The sample holder. The sample temperature is measured by attaching a Cernox temperature sensor to the upper surface of the sample (as shown in the right figure). The sensor is affixed to a copper plate, which is then pressed against the sample using a Belleville spring. (Note: the orientation is opposite to that of Fig \ref{fig:3D_configuration_system}, with the top and bottom reversed.)} 
\end{figure}

We used a helium circulation-type mechanical refrigerator (model PS11SS; Nagase Techno-Engineering Co.) to cool the sample for refractive index measurement to cryogenic temperatures. This system is based on a 4~K-class GM cooler (Sumitomo Heavy Industries, RDK-101D) and provides the first-stage cooling capacity of 100 mW and the second-stage cooling capacity of 3 W. This allows the sample (sample holder) placed on the cold work surface with a diameter of 3 cm to be cooled to below 5 K in approximately 2.5 hours under no-load conditions. The cold work surface is equipped with a heater and a Si temperature sensor, providing temperature control functionality. The temperature of the cold work surface can be controlled to any desired temperature up to approximately 200 K with the stability of $\pm$0.2 K (Fig \ref{fig:sample_holder} left). Hereafter, the temperature of the cold work surface measured by the Si sensor is referred to as the ``Cold head temperature.''

The appearance of the sample holder is shown in Fig \ref{fig:sample_holder}. The holder is made of oxygen-free copper (OFHC, RRR100), which has a high thermal conductivity $>678 \mathrm{W \, m^{-1} \, K^{-1}}$ at 4 -- 70~K. To ensure good thermal contact between the prism and the cold work surface via the sample holder, a Cernox temperature sensor plate is pressed by a Belleville spring. Hereafter, the temperature measured by the Cernox sensor on this plate is referred to as the ``Sample temperature.'' This is because the Cernox sensor is in contact with the sample, allowing it to measure the sample's temperature. The sample holder and prism contract as they cool to cryogenic temperatures. The reason for using a Belleville spring is to provide sufficient spring contraction, greater than the contraction of the sample holder when cooled from room temperature to cryogenic temperatures, in order to maintain good thermal contact between the prism and holder (Fig \ref{fig:sample_holder} right). The pressure at the contact surface is approximately 0.70 MPa. We assume the fracture strength of CdZnTe is similar to CdTe, which is estimated to be 22 MPa \cite{AZoM_CdTe}.
Based on this assumption, this design provides a significant safety margin.

In practice, however, there is external heat input from the room-temperature environment through vacuum windows and thermal contact input from the temperature sensor leads. As a result, the lowest attainable temperature of the refractive index measurement sample is 12.57 $\pm$ 0.14 K (See section~\ref{Measurements and Results}).

\section{Measurements and Results}
\label{Measurements and Results}
To calculate $n$, the values of $\delta_{\text{min}}$ and $\alpha$ are required, as shown in Equation \ref{eq:minimum_deviation_method}. In these measurements, we determined $n$ of CdZnTe at a wavelength of 10.68 $\mu$m from cryogenic temperature to room temperatures by following the procedure outlined below:

\begin{enumerate}
    \item Angle calibration measurement
    \item Apex angle measurement at room temperature
    \item Deviation angle measurement from cryogenic temperatures to room temperature
\end{enumerate}

The following sections explain the procedure of each measurement and its data analysis.

\subsection{Angle calibration measurement}
\label{Angle calibration measurement}
The angle calibration measurement determines the reference point of the movable stage position corresponding to the origin of the angle measurement. When the rotating sample stage is empty, the collimated measurement beam exiting from the fixed stage passes straight through the rotation center of the rotating sample stage. When the movable stage rotates, it captures the collimated measurement beam on its focusing system. 
The Si PD and MCT PV detectors output voltage, which converts incident infrared light strength into an electrical signal, varies with the rotation angle of the movable stage \( \theta_{\text{msta}} \) and forms a Gaussian-like profile as a function of $\theta$. The MCT detector and PD mentioned in section~\ref{Optical system} are used for this calibration measurement in the infrared and visible wavelengths, respectively.
The exact angle of a given beam is determined by fitting the obtained measurement profile with a Gaussian. The peak value of the Gaussian is defined as the calibration angle $\theta_0$ for $\alpha$ and the deviation angle measurements. This procedure is applied to subsequent angle measurements as well.

Table~\ref{table:calibration} shows the results of the angle calibration measurements with visible and infrared light. Using the results of visible light, which allows easier alignment, as a reference, it is found that the results for infrared light are consistent within 0.4\%. Furthermore, no significant difference is observed in the angle calibration measurements with and without the vacuum windows of shroud for infrared light.
The calibration angle $\theta_0$ varies for each measurement due to recalibration. The visible light measurement results are used for the apex angle calibration because the apex angle measurement is conducted at a wavelength of 635 nm. The infrared light measurement results with the windows are used for the minimum deviation angle calibration because the windows is necessary for measuring $n$ at low temperatures.

\begin{table}[ht]
\caption{Calibration angle $\theta_0$ in visible and infrared light measurements}
\label{table:calibration}
\begin{center}       
\begin{tabular}{|l|l|} 
\hline
\rule[-1ex]{0pt}{3.5ex}  Light Source & Calibration angle $\theta_0$ (degree)  \\
\hline\hline
\rule[-1ex]{0pt}{3.5ex}  Visible light source at 635 nm & $23.415 \pm 0.001$  \\
\hline
\rule[-1ex]{0pt}{3.5ex}  Infrared light source at 10.68 $\mu$m & $23.338 \pm 0.003$   \\
\hline
\rule[-1ex]{0pt}{3.5ex}  Infrared light source at 10.68 $\mu$m with windows & $23.340 \pm 0.007$   \\
\hline
\end{tabular}
\end{center}
\end{table}

\subsection{Apex angle measurement}
The apex angle \( \alpha \) of the CdZnTe sample is measured using autocollimation measurements on the two surfaces (Surface A and Surface B) adjacent to $\alpha$.  
The autocollimation measurement is a high-precision angular measurement technique that utilizes the reflection of light \cite{born_wolf_optics}. This measurement is conducted at room temperature and atmospheric pressure.  
Each autocollimation measurement is performed three times.


\begin{figure}[H]
\begin{center}
\begin{tabular}{c}
\includegraphics[height=9.0cm]{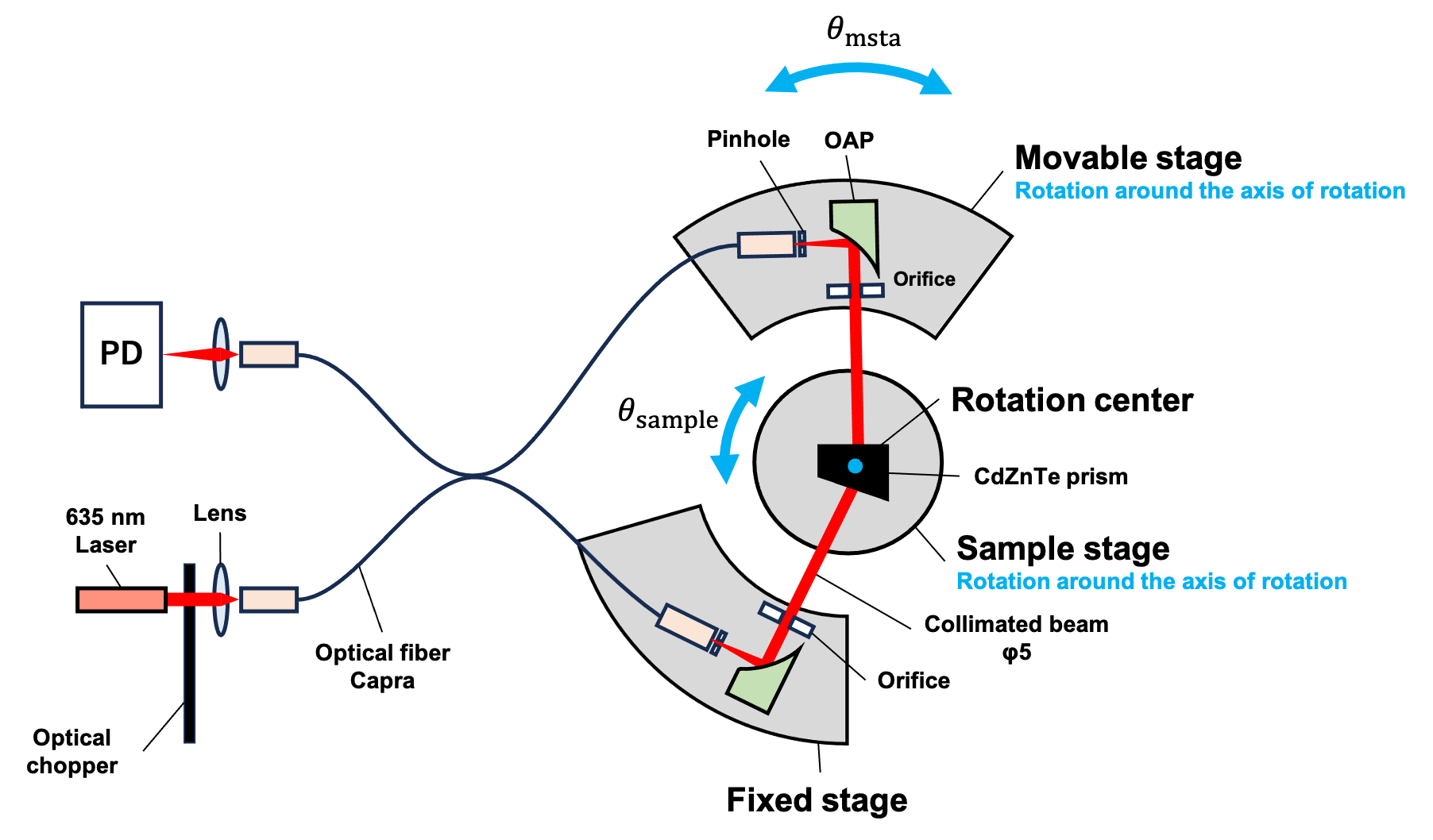}
\end{tabular}
\end{center}
\caption 
{ \label{fig:autocollimation_setup}
A schematic diagram of the apex angle measurement. The arrows represent the direction of the measurement light. The red one indicates the light traveling from the light source toward the surface of the prism, and the blue one represents the light reflected off the surface of the prism and traveling toward the detector.} 
\end{figure}

Fig \ref{fig:autocollimation_setup} shows a schematic diagram of the apex angle measurement setup. 
As described in Section~\ref{Angle calibration measurement}, both the sample stage and the movable stage rotate around the central rotation axis of the sample stage.  
In this measurement, the prism is placed at the center of the sample stage. Light from a visible laser source is focused by a lens and enters the optical fiber coupler. This fiber is used instead of a beam splitter to simplify the autocollimation measurement. The light entering the fiber coupler is split into a 50:50 ratio at the branching point of the coupler. One path is transmitted to the fiber end on the fixed stage, while the other is transmitted to the fiber end on the movable stage. The light exits each fiber end as diffused light, which is then reflected by an OAP mirror to form a collimated beam with a diameter of 10 mm.

\begin{figure}[H]
\begin{center}
\begin{tabular}{c}
\includegraphics[height=6.0cm]{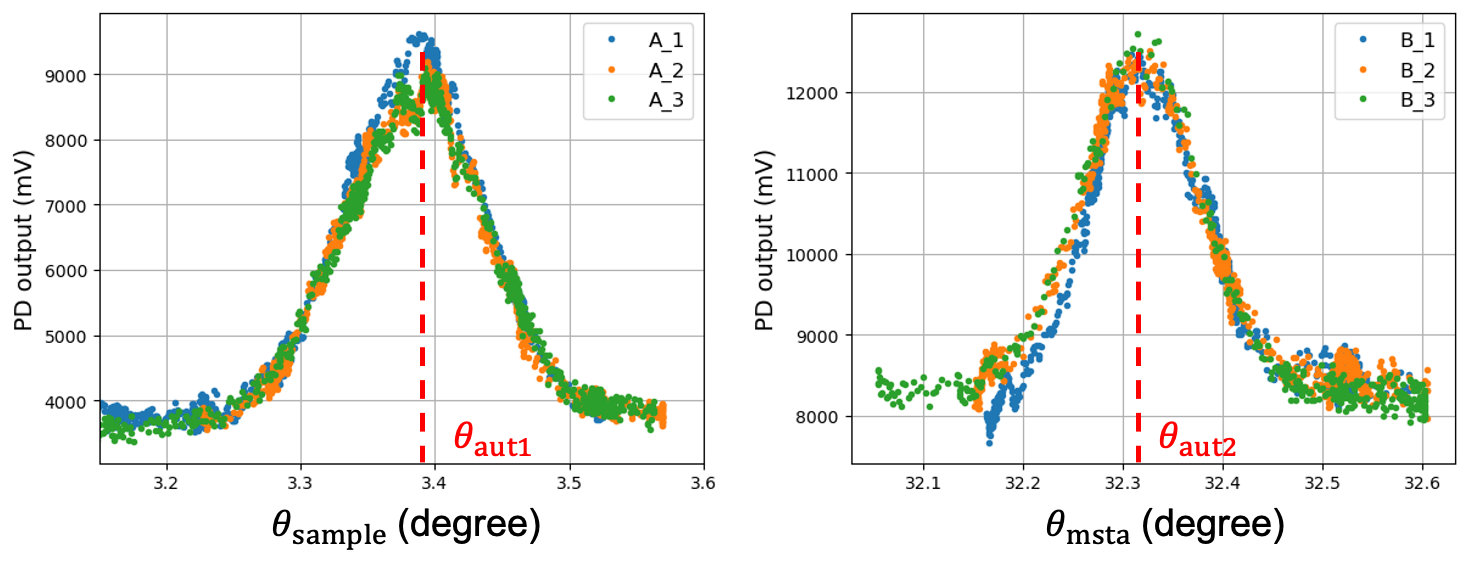}
\end{tabular}
\end{center}
\caption 
{ \label{fig:autocoli a&b}
The autocollimation measurement results for Surface A (left panel) and Surface B (right panel) of the sample.  
Each measurement is repeated three times.  
The peak values for each measurement are fitted with a Gaussian function, and the autocollimation angles \( \theta_{\text{aut1}} \) and \( \theta_{\text{aut2}} \) are determined.}
\end{figure} 

First, we describe the autocollimation measurement on the fixed stage.  
The collimated beam produced by the OAP mirror is incident on Surface A of the prism and is reflected.  
As the rotation angle of the sample stage \( \theta_{\text{sample}} \) is gradually rotated, the collimated beam striking Surface A is precisely reflected back to the OAP mirror.  
The OAP mirror then reflects the light back into the optical fiber coupler, where it is transmitted through the fiber and exits as diffused light at the opposite fiber end.  
This light is refocused by a lens and detected by the PD.  
Due to the characteristics of the optical system, the detector voltage as a function of \( \theta_{\text{sample}} \) follows a Gaussian-like distribution as shown in Fig~\ref{fig:autocoli a&b}~(left).   
By fitting the obtained measurement profile with a Gaussian function, the peak value is determined.  
The corresponding angle is defined as the autocollimation angle \( \theta_{\text{aut1}} \).

Next, we describe the autocollimation measurement on the movable stage.  
The rotation angle of the sample stage \( \theta_{\text{sample}} \) is set to \( \theta_{\text{aut1}} \) and fixed.  
As in the measurement for Surface A, the collimated beam produced by the OAP mirror is incident on Surface B of the prism and is reflected.  
As the rotation angle of movable stage \( \theta_{\text{msta}} \) is gradually rotated, the collimated beam striking Surface B is precisely reflected back to the OAP mirror.  
The OAP mirror then reflects the light back into the optical fiber coupler, and the transmitted light is detected by the PD in the same manner as before.  
The autocollimation angle for Surface B is defined as \( \theta_{\text{aut2}} \) as shown in Fig~\ref{fig:autocoli a&b}~(right). By using the optical fiber coupler, the collimated beam can be generated from the OAPs on both sides. This simplified the optical system.

Finally, the apex angle \( \alpha \) which is the difference between \( \theta_{\text{aut1}} \) and \( \theta_{\text{aut2}} \) is obtained by

\begin{equation}
\alpha = \theta_{\text{aut2}} - \theta_{\text{0}}.
\label{eq:alpha}
\end{equation}

As a result, $\alpha$ of the prism is found to be $30.033 \pm 0.001^\circ$. In the current derivation of the $n$, we assumed that $\alpha$ does not have temperature dependence. This is because the crystal structure of CdZnTe is cubic, and we assume that the contraction due to cooling is isotropic. If there is any change, it is expected to affect the determination of $n$ as well in a fashion formulated in the following subsection.

\subsection{Deviation angle measurement}
In this study, the deviation angle \( \delta \) is measured for each \( \theta_1 \), and the minimum deviation angle \( \delta_{\text{min}} \) is obtained from the dependence of \( \delta \). Each measurement is repeated multiple times.


\begin{figure}[H]
\begin{center}
\begin{tabular}{c}
\includegraphics[height=9.0cm]{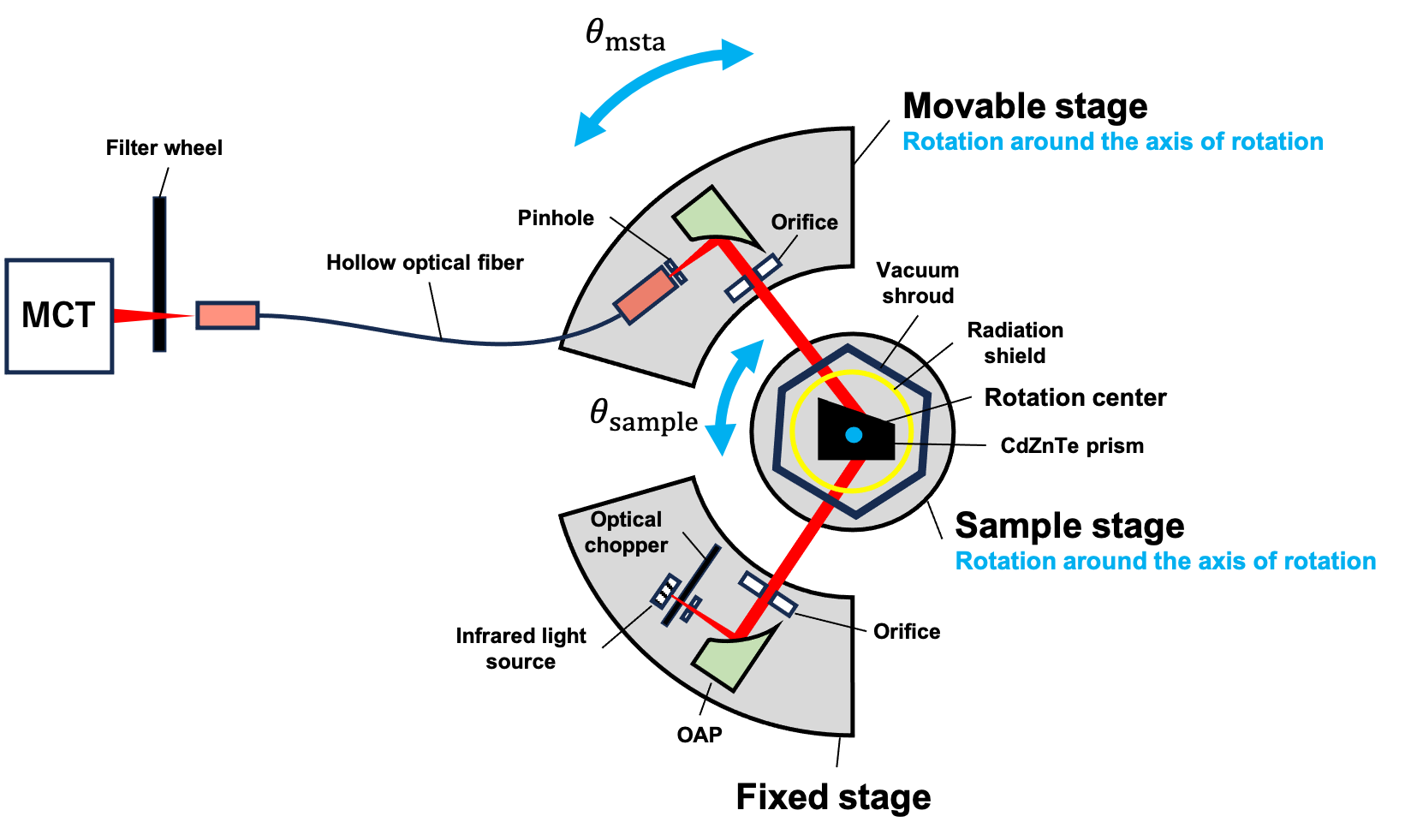}
\end{tabular}
\end{center}
\caption 
{ \label{fig:deviation_setup}
A schematic diagram of the deviation angle measurement. The arrows represent the direction of the measurement light. The 3D diagram is shown in Fig \ref{fig:3D_configuration_system}.} 
\end{figure}

Fig \ref{fig:deviation_setup} shows a schematic diagram of the deviation angle measurement setup. 
The CdZnTe sample is placed at the center of the sample stage.  
A pinhole with a diameter of 0.5 mm is placed in front of the diffused light from the infrared light source to create a pseudo point source.  
The light is reflected by the OAP mirror, forming a collimated beam with a diameter of 10 mm.
This collimated beam passes through the vacuum shroud window and the radiation shield hole, refracts through the CdZnTe sample, and exits through the radiation shield hole and the vacuum shroud window.   
The movable stage is rotated to capture the collimated beam with the OAP mirror. 
The light reflected by the OAP mirror enters a hollow optical fiber and is transmitted to the MCT detector.
Due to the characteristics of the optical system, the detector voltage as a function of \( \theta_{\text{sample}} \) follows a Gaussian-like distribution.  
The hollow optical fiber used in this measurement has an aperture diameter of 1 mm. The fiber is designed specifically for a wavelength of 10.68 $\mu$m and is manufactured by covering the inner wall of a glass tube having an inner diameter of 1 mm with thin films coated with silver (Ag) and silver iodide (AgI)\cite{Matsuura_1995}. 
To restrict the width of the angular profile, a 0.5 mm diameter pinhole is placed in front of the fiber aperture. To improve the accuracy of the Gaussian peak determination, we perform this process. A BPF of 10.68~$\mu$m is placed in front of the MCT detector.

As illustrated in Fig~\ref{fig:prism_refraction}, the relation between \( \alpha \) of the prism and the refracted angles \( \theta'_1 \) and \( \theta'_2 \) becomes


\begin{equation}
\alpha = \theta'_1 + \theta'_2.
\label{eq:geometry}
\end{equation}

On the other hand, the Snell's law at the incident and exit surfaces can be written as
\begin{equation}
n_0 \sin \theta_1 = n \sin \theta'_1 \quad \text{and} \quad n_0 \sin \theta_2 = n \sin \theta'_2.
\label{eq:snell_law}
\end{equation}

Combining the Equations \ref{eq:geometry} and \ref{eq:snell_law}, we obtain
\begin{equation}
\delta(\theta_1) = \theta_1 + \theta_2 - \alpha \\
= \theta_1 - \alpha + \arcsin \left[ n \sin \left( \alpha - \arcsin \left( \frac{\sin \theta_1}{n} \right) \right) \right].
\label{eq:fitting_function}
\end{equation}
Here, the external medium is assuemed to be vacuum ($n_0 = 1$).

\begin{figure}[H]
\begin{center}
\begin{tabular}{c}
\includegraphics[height=10.5cm]{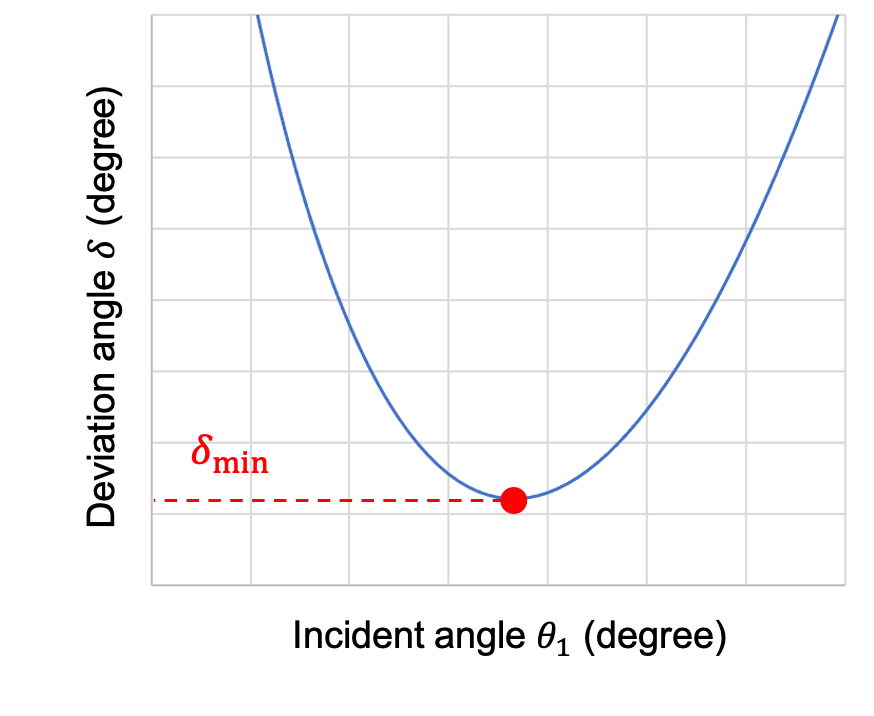}
\end{tabular}
\end{center}
\caption 
{ \label{fig:deviation_theory}
Example of the relation between incidence angle $\theta_1$ and $\delta$ based on Equation~\ref{eq:fitting_function}}
\end{figure}

When Equation \ref{eq:fitting_function} is plotted with $\theta_1$ on the horizontal axis and $\delta$ on the vertical axis, a curve like the one shown in Fig \ref{fig:deviation_theory} is produced. The deviation angle $\delta$ reaches its minimum, 
$\delta_{\text{min}}$, when $\theta_1 = \theta_2$. In this study, the deviation angle $\delta$ is measured at various $\theta_1$, and the results are fitted with Equation \ref{eq:fitting_function} to determine $\delta_{\text{min}}$.

\begin{figure}[H]
\begin{center}
\begin{tabular}{c}
\includegraphics[height=9.5cm]{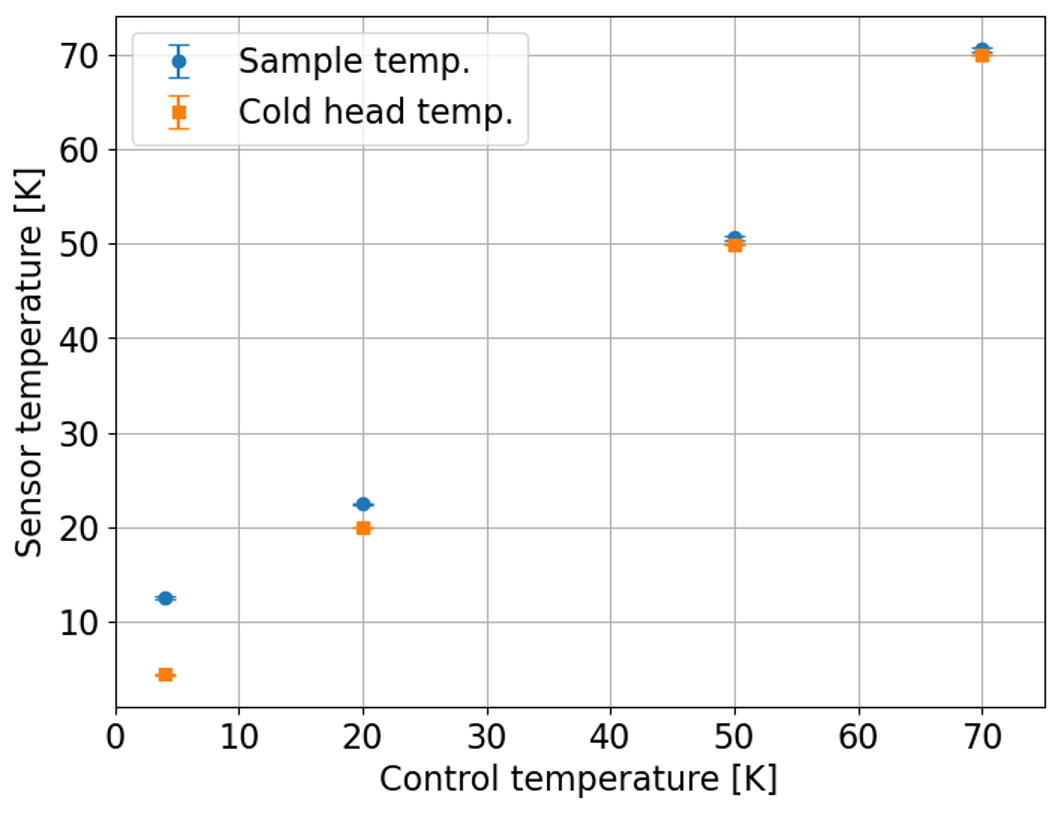}
\end{tabular}
\end{center}
\caption 
{ \label{fig:temp_sample}
Measured temperatures vs control temperature in this experiment. The sample temperature is stable within 0.1~K at 20~K and below; 0.25~K at 50, and 70~K (control temperatures), respectively.} 
\end{figure}

We measured $\delta$ at room temperature ($ 298 \pm 1$ K) and 4, 20, 50, and 70 K, controlled by the heater which is built inside the cold head. Fig \ref{fig:temp_sample} shows the cold head temperature and sample temperature at each controlled temperature during the deviation angle measurements. Table \ref{table:temp_sample} summarizes the temperatures for each controlled temperature of 4, 20, 50, and 70 K.

The temperature difference between the cold head temperature and the sample temperature is attributed to external heat input from the room temperature environment through the vacuum windows, as well as thermal contact input from the Cernox temperature sensor leads measuring the sample temperature. We believe that the latter can be minimized in future setups.

\begin{table}[H]
\caption{Measured temperatures of the cold head and sample during refractive index measurements.}
\label{table:temp_sample}
\begin{center}       
\begin{tabular}{|l|l|l|l|} 
\hline
\rule[-1ex]{0pt}{3.5ex}  Control temp. (K) & Cold head temp. (K) & Sample temp. (K)  \\
\hline\hline
\rule[-1ex]{0pt}{3.5ex}  70 & $69.97 \pm 0.05$ & $70.57 \pm 0.23$  \\
\hline
\rule[-1ex]{0pt}{3.5ex}  50 & $49.94 \pm 0.05$ & $50.59 \pm 0.20$  \\
\hline
\rule[-1ex]{0pt}{3.5ex}  20 & $19.99 \pm 0.03$ & $22.47 \pm 0.06$  \\
\hline
\rule[-1ex]{0pt}{3.5ex}  4  & $4.43 \pm 0.09$ & $12.57 \pm 0.14$   \\
\hline
\end{tabular}
\end{center}
\end{table}

\begin{figure}[H]
\begin{center}
\begin{tabular}{c}
\includegraphics[height=8.5cm]{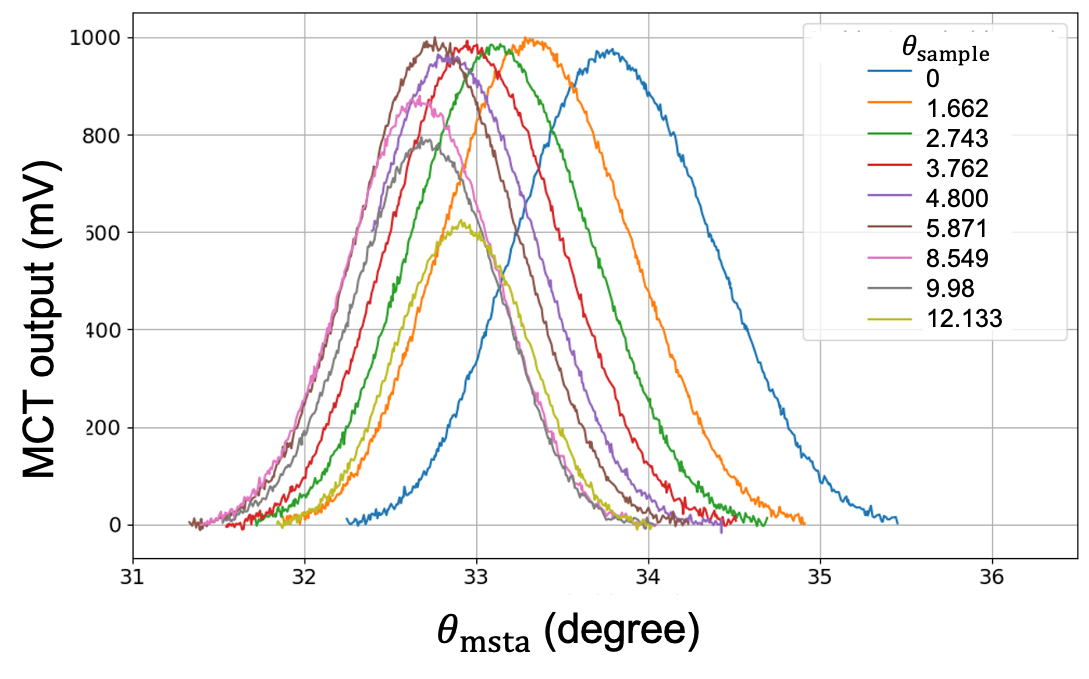}
\end{tabular}
\end{center}
\caption 
{ \label{fig:deviation_mesurement}
Results of deviation angle measurements at 22.47 $\pm$ 0.06 K for \( \theta_{\text{sample}} \) shown in the legend. The horizontal axis represents \( \theta_{\text{msta}} \),} and the vertical axis shows the MCT detector output at these angle. 
\end{figure}

Fig \ref{fig:deviation_mesurement} shows the results of deviation angle measurements at 22.47 $\pm$ 0.06 K for various $\theta_{\text{sample}}$. The horizontal axis represents the rotation angle of the movable stage $\theta_{\text{msta}}$, and the vertical axis shows the MCT output. The peak values of each measurement are fitted with a Gaussian to determine $\delta$. 
At each $\theta_{\text{sample}}$, the acquisition and fitting of the MCT output curve as a function of $\theta_{\text{msta}}$ is conducted three times to decrease statistical errors.
The same procedure is followed for the deviation angle measurements at other temperatures.


\begin{figure}[H]
\begin{center}
\begin{tabular}{c}
\includegraphics[height=15cm]{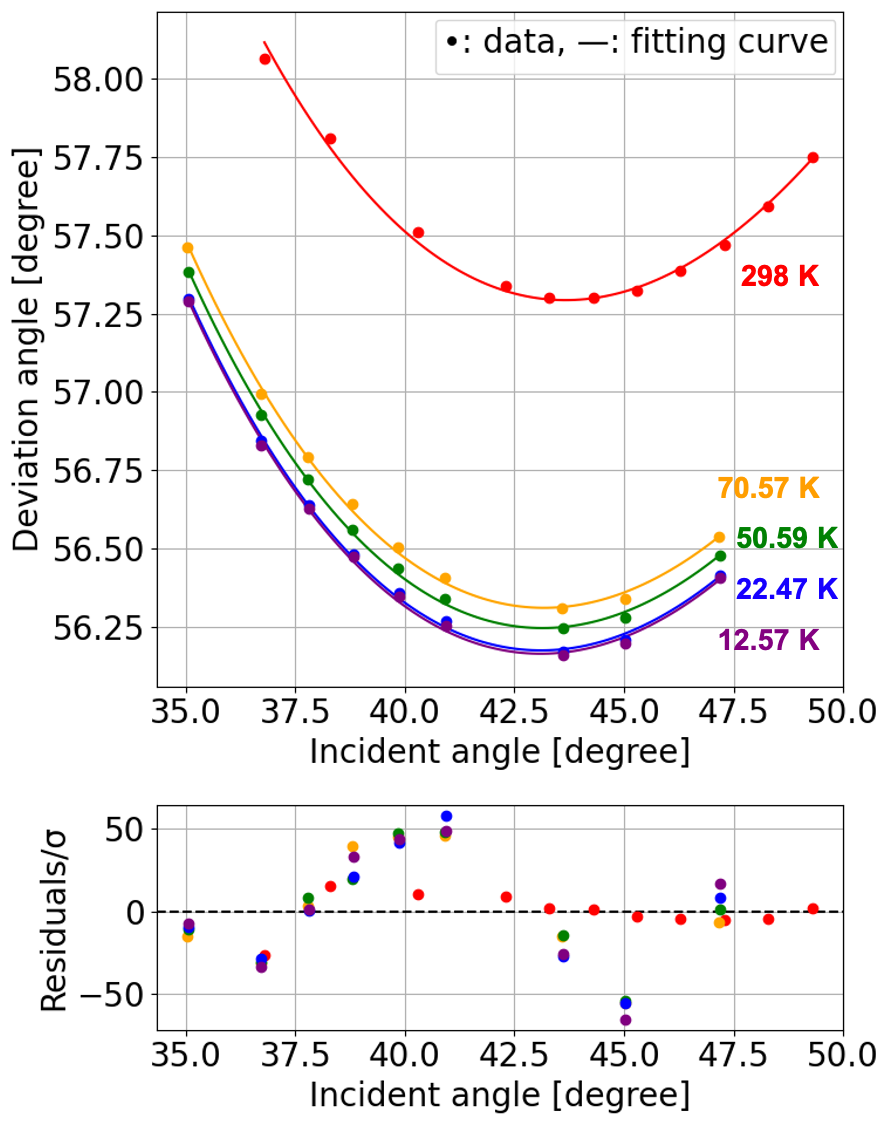}
\end{tabular}
\end{center}
\caption 
{ \label{fig:deviation_temp_result}
The upper panel shows the results of the deviation angle measurements of CdZnTe at each control temperature at a wavelength of 10.68 $\mu$m, along with the fitting curves based on Equation \ref{eq:fitting_function}.
The lower panel shows the residuals for each fitting. $\sigma$ is the fitting error obtained from the deviation angle measurements (e.g., Fig \ref{fig:deviation_mesurement}).} 
\end{figure}

The upper panel of Fig \ref{fig:deviation_temp_result} shows $\delta$ of the CdZnTe sample measured at 4, 20, 50, 70 K (controlled temperatures), and $ 298 \pm 1$ K at 10.68 $\mu$m. The vacuum pressure inside the vacuum shroud during cryogenic measurements is $(2.5 \pm 0.9) \times 10^{-3}$ Pa. 
The horizontal axis shows $\theta_{\text{msta}}$ and the vertical axis shows the MCT output. The peak values of each measurement's data are fitted with a Gaussian distribution to determine $\delta$. Since $\delta$ at each $\theta_{\text{sample}}$ is measured multiple times, the weighted average of $\delta$, obtained from each fitting is calculated. The fitting accuracy of $\delta$, using Gaussian fitting is approximately from 0.001° to 0.01°.

Fig~\ref{fig:deviation_temp_result} shows the results of the fitting using Equation~\ref{eq:fitting_function}. The colors of the symbols represent the control temperatures.
The solid lines represent the curves of Equation \ref{eq:fitting_function} fitted to the results at each temperature, using $n$ and \(\theta_{\text{offset}}\) as a free parameter.
The incident angle \(\theta_1\) on the horizontal axis of Fig~\ref{fig:deviation_temp_result} is corrected by \(\theta_{\text{offset}}\).

\begin{figure}[H]
\begin{center}
\begin{tabular}{c}
\includegraphics[height=10cm]{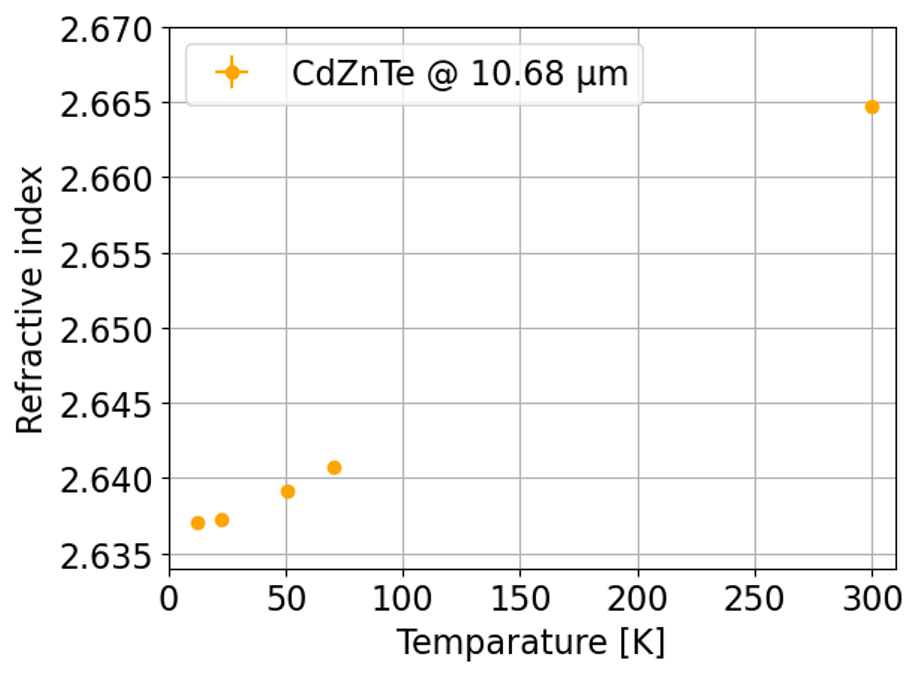}
\end{tabular}
\end{center}
\caption 
{ \label{fig:temp_refractive-index}
Temperature dependence of $n$ of CdZnTe from cryogenic to room temperature. Only the statistical error is reflected in the plot. Error bars are smaller than the symbols.}
\end{figure}

The refractive index $n$ of CdZnTe at 10.68 $\mu$m is determined, as shown in Table \ref{table:temp_refractive-index}. And Fig~\ref{fig:temp_refractive-index} illustrates the temperature dependence of $n$ of CdZnTe at 10.68 $\mu$m.
The lower panel of Fig \ref{fig:deviation_temp_result} shows the residual divided by the statistical error $\sigma$ at each data point for each fitting. The residuals are much larger than $\sigma$, and show the presence of a systematic pattern. This suggests that unintended effects occurred during the actual deviation angle measurements. This also indicates that the model in Equation \ref{eq:fitting_function} may be insufficient.  The next section~\ref{Measurement Uncertainties} discusses the uncertainties in angle measurements for this experimental setup.

\begin{table}[ht]
\caption{Temperature dependence of $n$ of CdZnTe at a wavelength of 10.68 $\mu$m.} 
\label{table:temp_refractive-index}
\begin{center}       
\begin{tabular}{|l|l|} 
\hline
\rule[-1ex]{0pt}{3.5ex}  Sample temp. [K] & Refractive index  \\
\hline\hline
\rule[-1ex]{0pt}{3.5ex}  $298 \pm 1$ & $2.6647 \pm 0.0007$  \\
\hline
\rule[-1ex]{0pt}{3.5ex}  $70.57 \pm 0.23$ & $2.6407 \pm 0.0022$  \\
\hline
\rule[-1ex]{0pt}{3.5ex}  $50.59 \pm 0.20$ & $2.6391 \pm 0.0022$  \\
\hline
\rule[-1ex]{0pt}{3.5ex}  $22.47 \pm 0.06$ & $2.6373 \pm 0.0022$  \\
\hline
\rule[-1ex]{0pt}{3.5ex}  $12.57 \pm 0.14$ & $2.6371 \pm 0.0022$  \\
\hline
\end{tabular}
\end{center}
\end{table}

\section{Measurement Uncertainties}
\label{Measurement Uncertainties}

There are various effects in this system that are not considered in Equation \ref{eq:fitting_function}, and these effects could potentially contribute to the uncertainties in the angle measurements. Table \ref{table:Uncertainties} shows the different factors contributing to angular errors and their worst-case estimates. Below, the details of the evaluation for each factor are described.

\begin{figure}[H]
\begin{center}
\begin{tabular}{c}
\includegraphics[height=8cm]{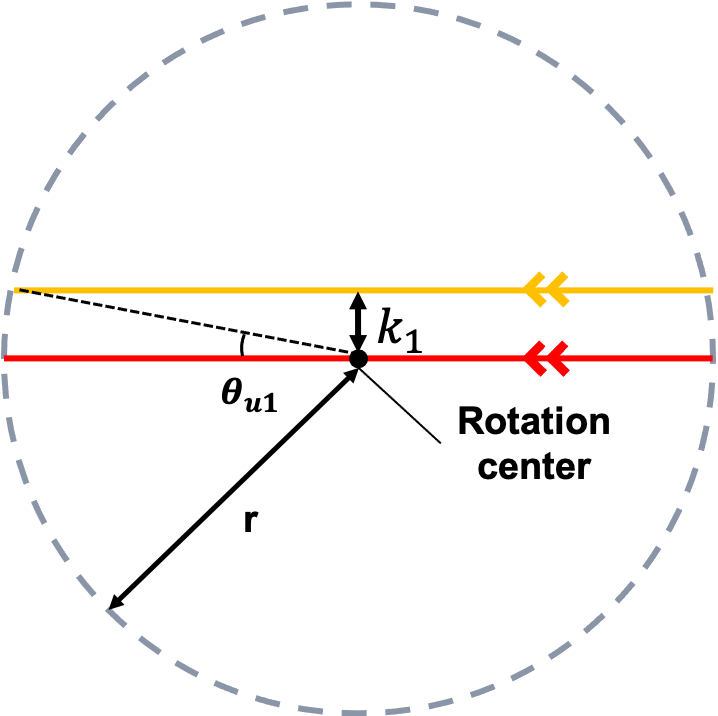}
\end{tabular}
\end{center}
\caption 
{ \label{fig:Uncertainties_1}
The schematic diagram shows the angular error due to the lateral shift of the collimated measurement beam. The red line represents the collimated measurement beam passing through the center, while the orange line represents the collimated measurement beam that is shifted laterally. The gray dashed line represents the rotation of the movable stage.} 
\end{figure}

First, we discuss the angular error \( \theta_{\textnormal{u1}} \) due to the lateral shift of the collimated measurement beam (Fig \ref{fig:Uncertainties_1}). The collimated measurement beam needs to pass through the axis of rotation of the measuring device. We placed an orifice with a diameter of 0.5 mm along the axis of rotation. The visible light collimated measurement beam is aligned so that it passed through this orifice. The precision of the visual confirmation, denoted as \( k_1 \), is estimated to be within a quarter of the orifice size, \( k_1 < 0.125 \, \text{mm} \), under the worst-case assumption. The corresponding angular error is estimated as \( \theta_{\textnormal{u1}} \leq 0.033^\circ \) by 

\begin{equation}
\sin \theta_{\textnormal{u1}} = \frac{k_1}{r}.
\label{eq:Uncertainties_1}
\end{equation}
The $r$ is the rotation radius of the movable stage.

\begin{figure}[H]
\begin{center}
\begin{tabular}{c}
\includegraphics[height=8cm]{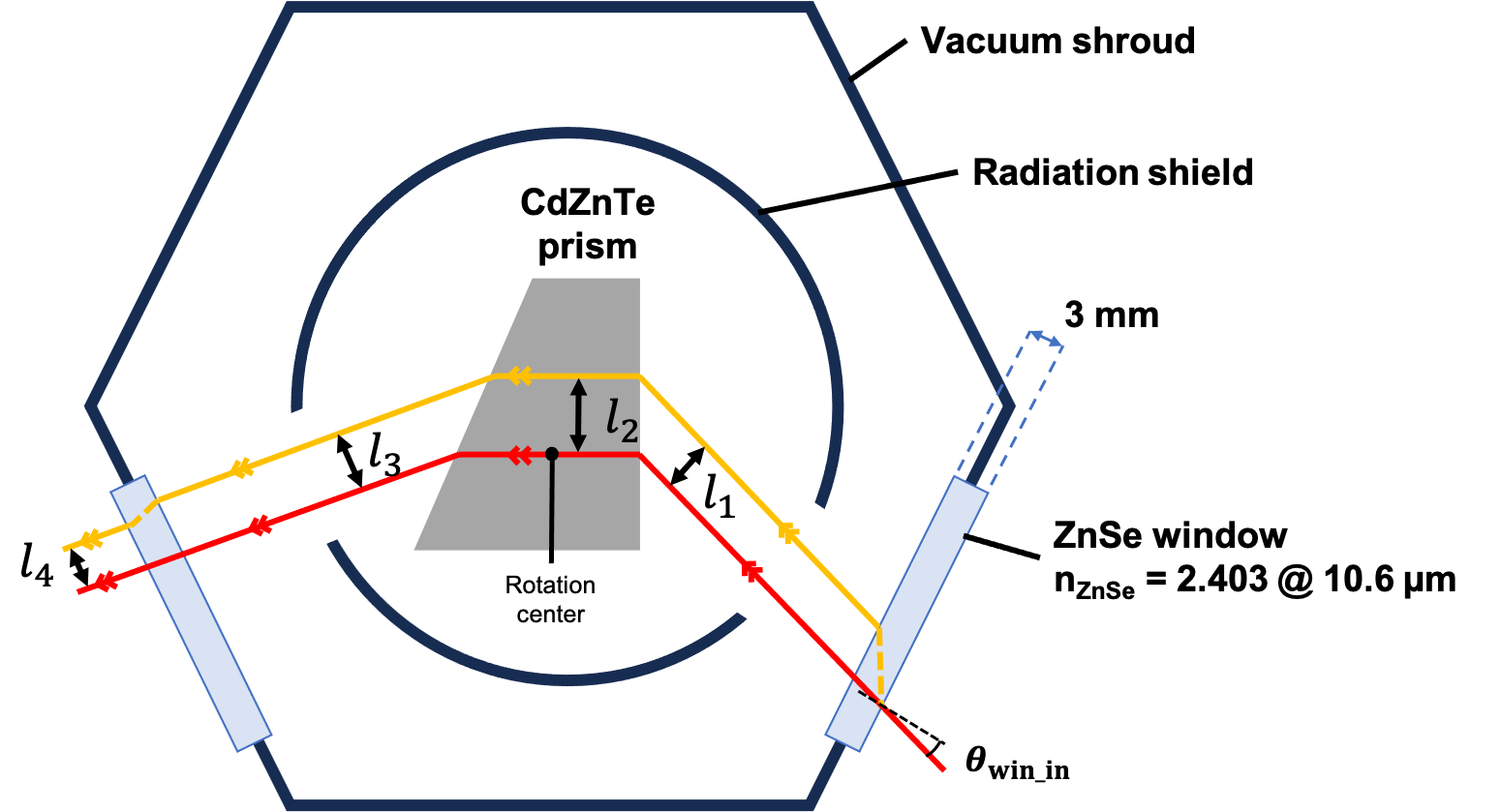}
\end{tabular}
\end{center}
\caption 
{ \label{fig:Uncertainties_2}
The schematic diagram shows the amount of shift in the collimated measurement beam due to the ZnSe window. The red line represents the collimated measurement beam when there is no window, while the orange line represents the beam that is shifted due to the presence of the ZnSe window.} 
\end{figure} 

Next, we describe the angular error $\theta_{\textnormal{u2}}$ caused by the lateral displacement of the collimated measurement beam due to the vacuum windows. As shown in Fig \ref{fig:Uncertainties_2}, two 3-mm thick ZnSe windows ($n_{\textnormal{ZnSe}} = 2.3950$ at 10.6 $\mu$m \cite{Querry_1987}) are used on both the incident and exit sides. The angular error $\theta_{\textnormal{u2}}$ is a function of $\theta_{\textnormal{win\_in}}$ on the vacuum windows. In this measurement, the incident angle $\theta_1$ ranged from 35.0 to 47.5 degrees. Using Snell's law and the geometric relationships, we derive $l_4$, the displacement distance of the collimated exit beam caused by the effect of the window. The angular error is estimated to be $\theta_{\textnormal{u2}} < 0.06^\circ$ by

\begin{equation}
\theta_{u2} = \tan^{-1} \left( \frac{l_4}{r} \right).
\label{eq:Uncertainties_2}
\end{equation}

The angular measurement error caused by the encoder, treated as a random error, is $\theta_{\textnormal{u3}} <  6 \times 10^{-4}$ degree. The encoder used is the Heidenhain RON806, and $\theta_{\textnormal{u3}}$ is based on its specifications.

Finally, we explain the angular error $\theta_{\textnormal{u4}}$ caused by the tilt of the sample around the vertical axis. When a tilt $\theta_{tilt}$ occurs, the apex angle $\alpha$ changes to $\alpha’$ as shown in Equation \ref{eq:Uncertainties_4}. Since a change in $\alpha$ affects $\delta$, it introduces an angular error as

\begin{equation}
\alpha' = \frac{\alpha}{\cos \theta_{\text{tilt}}}.
\label{eq:Uncertainties_4}
\end{equation}
When the $\theta_{\textnormal{tilt}}$ is less than or equal to 1 degree, the refractive index error is evaluated as $\Delta n \leq 10^{-4}$ based on Equation \ref{eq:minimum_deviation_method}. In practice, we confirmed that the tilt is less than 1 degree using an auto-collimation measurement.

The time stability of the light source and detector is evaluated as part of the statistical error by performing multiple measurements in this experiment. Therefore, the total uncertainties that should be added to the refractive index error amount to $\theta_{\textnormal{uall}} < 0.093^\circ$. Here, the total uncertainty is linearly summed. The systematic uncertainty is dominated by $\theta_{\textnormal{u2}}$. It is estimated at the maximum value over the entire range of $\theta_1$ to the sample. Considering that $\delta_{\text{min}}$ is determined from data at smaller incident angles, the angular error $\theta_{\textnormal{u2}}$ is likely an overestimation. However, in this paper, the uncertainty in $n$, calculated from the total uncertainties $\theta_{\textnormal{uall}}$, is presented as $\Delta n_{\text{total}} < 2.2 \times 10^{-3}$ at the temperature range from 12.57 K to 70.57 K, and $\Delta n = \pm 0.0007$ at 298~K.

To verify the reliability of the absolute values of $n$ measured with our setup, we performed a test measurement using a prism of CaF$_2$, whose $n$ at room temperature at a wavelength of 635 nm is well known. As a result, the refractive index $n$ obtained in this system is found to be consistent with Yamamuro $et$ $al.$ 2006 \cite{Yamamuro_2006} within the reported uncertainty of $3 \times 10^{-4}$. Therefore, our measurement system and procedure are reliable at least to some extent. However, the uncertainty in $n$ of CdZnTe obtained here is twice larger than the accuracy requirement of $\Delta n < 10^{-3}$. The current estimation of uncertainties seems to be an overestimation, and the presence of systematic error is suggested by the systematic pattern in the fitting residuals (Fig \ref{fig:deviation_temp_result} bottom). Further investigation of the error causes and improvement of the system are needed to achieve $\Delta n < 10^{-3}$.

Although not included in the measurement uncertainties, the refractive index $n$ may vary within a single CdZnTe crystal. The CdZnTe measured in this study has low impurity levels, so the non-uniformity is considered to be small, and we assume that its impact on $n$ is not significant.

\begin{table}[H]
\caption{List of the measurement uncertainties.} 
\label{table:Uncertainties}
\begin{center}       
\begin{tabular}{|l|l|p{9cm}|} 
\hline
\rule[-1ex]{0pt}{3.5ex}  Title & Estimated error (degree) & Description  \\
\hline\hline
\rule[-1ex]{0pt}{3.5ex}   $\theta_{\textnormal{u1}}$ & $ < 3.3 \times 10^{-2}$ & Angular error due to the lateral displacement of the measurement beam ($\phi 5$) from the rotation axis of the measurement system  \\
\hline
\rule[-1ex]{0pt}{3.5ex}   $\theta_{\textnormal{u2}}$ & $< 6 \times 10^{-2}$ & Angular error due to lateral displacement of the measurement beam caused by the vacuum windows, 3 mm thick ZnSe window with high refractive index ($n = 2.403$ @ 10.6 $\mu$m)  \\
\hline
\rule[-1ex]{0pt}{3.5ex}   $\theta_{\textnormal{u3}}$ & $ < 6 \times 10^{-4}$ & Specification of the encoder error \\
\hline
\rule[-1ex]{0pt}{3.5ex}   $\theta_{\textnormal{u4}}$ & $ < 1.0 \times 10^{-4}$ & Angular error caused by an increase in the distance the collimated measurement beam passes through due to the tilt between the prism surfaces \\
\hline
\rule[-1ex]{0pt}{3.5ex}  \textbf{Total} & $ < 9 \times 10^{-2}$ & \\
\hline
\end{tabular}
\end{center}
\end{table}

\section{Discussion}
\label{Discussion}

\begin{figure}[H]
\begin{center}
\begin{tabular}{c}
\includegraphics[height=12cm]{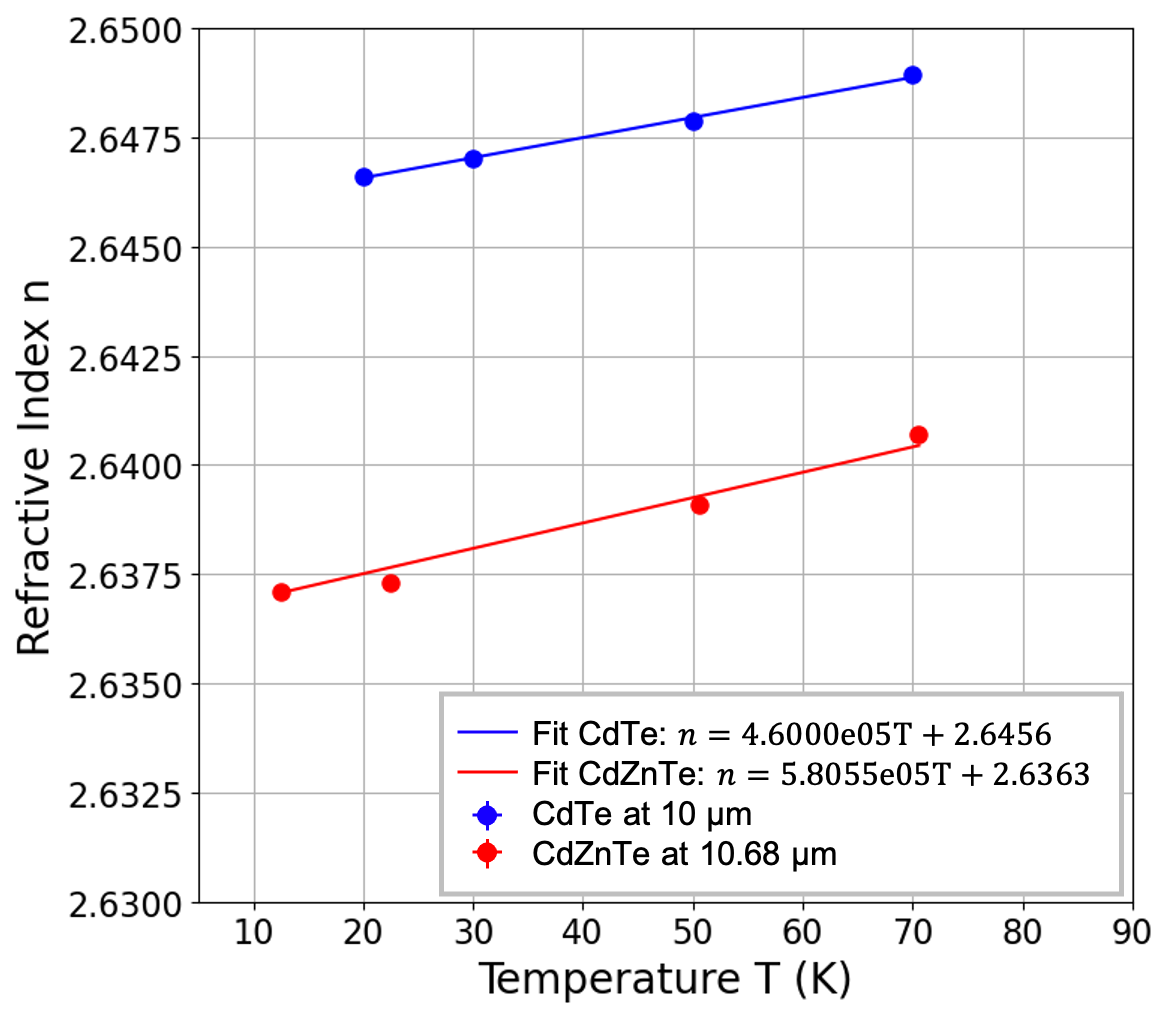}
\end{tabular}
\end{center}
\caption 
{ \label{fig:temp_refractive-index_fit}
Temperature dependence of $n$ of CdZnTe (green dots, this work). and CdTe (blue dots)~\cite{Browder91}. 
Fitting is performed considering only the statistical error. Error bars are smaller than the symbols.}
\end{figure}

The refractive index $n$ of CdZnTe in the mid-infrared region at cryogenic and room temperatures has not been directly measured before. Semiconductor materials like CdZnTe are known to often have differences in optical properties among batches, due to factors like the amount of doped impurities. So, it is important to check the changes in $n$ caused by differences among batches and the absence of birefringence. The composition formula of this sample is Cd$_{0.96}$Zn$_{0.04}$Te.

In the design of GREX-PLUS/HRS and its predecessor SPICA/SMI-HR, the refractive index $n$ of CdZnTe was estimated from that of CdTe \cite{Hlidek_2001}. Fig \ref{fig:temp_refractive-index_fit} shows the temperature dependence of the refractive indices of CdZnTe and CdTe. 
Linear fitting is performed on the refractive indices $n$ of CdZnTe and CdTe at \(T < 70 \, \text{K}\). The temperature dependence \(\Delta n / \Delta T\), corresponding to the slope of the line, is determined for each material. The temperature dependence \(\Delta n / \Delta T\) of CdZnTe at 10.68 \(\mu\)m is estimated be \((5.8 \pm 0.3) \times 10^{-5} \, \text{K}^{-1}\), while \(\Delta n / \Delta T\) of CdTe at 10 \(\mu\)m is \(4.6  \times 10^{-5} \, \text{K}^{-1}\) \cite{Browder91}.
We have assumed that there would not be significant qualitative differences between the physical properties of CdZnTe and CdTe and have proceeded with the initial design of the spectrometer based on this assumption. The current results are consistent with this prediction, providing validity to our initial design considerations. Moreover, the average $\Delta n/\Delta T$ between 12.57 $\pm$ 0.14 K and 298 $\pm$ 1 K is $(9.92 \pm 0.07) \times 10^{-5}$ K$^{-1}$ for CdZnTe at 10.68 $\mu$m and  $9.5 \times 10^{-5}$ K$^{-1}$ for CdTe at 10 $\mu$m \cite{Browder91}, and the two values are in fairly good agreement.

Using the system developed in this study, the refractive index $n$ of CdZnTe is measured with uncertainty of $\Delta n_{\text{total}} < 2.2 \times 10^{-3}$ over the temperature range from cryogenic temperatures to room temperature. It is expected that the system can be used to perform similar measurements at wavelengths other than 10.68 $\mu$m. 
The decrease in diffraction efficiency must be limited to 10\% so that the 17.754~$\mu$m H$2$O gas emission line remains within 0.18$\Delta \lambda$ around $\lambda_{\text{bla}}$. This requirement applies to GREX-PLUS/HRS, which is currently under development.
The same measurement of $n$ of CdZnTe at T $<$ 20K as in this paper should be conducted at around at the wavelength of the target line.
Measurements at the wavelength of the target line, 17.754~$\mu$m, are not performed because a suitable laser source is not available. The objective is to measure $n$ near this wavelength, but this is difficult due to the low optical strength at the wavelength.
Therefore, we first performed measurements at 10.68~$\mu$m to establish measurements in the mid-infrared. In the future, we plan to measure $n$ near the target wavelength using the same technique as the 10.68~$\mu$m measurement, using a BPF of 17~$\mu$m.

\section{Conclusion}
\label{conclusion}
In this study, we precisely measur the temperature dependence of $n$ of CdZnTe, a candidate material for the immersion grating to be installed on next-generation infrared space telescopes. Using the minimum deviation method, the refractive index $n$ of CdZnTe at a wavelength of 10.68 $\mu$m has been measured at controlled temperatures of 4, 20, 50, 70, and $ 298 \pm 1$ K. 
The measurement results of \( n \) are shown in the Table~\ref{table:temp_refractive-index}. The measurement accuracy is estimated as $\Delta n_{\text{total}} < 2.2 \times 10^{-3}$. From these results, the temperature dependence \(\Delta n / \Delta T\) of CdZnTe has been determined. These results also indicate that \(\Delta n / \Delta T\) of CdZnTe and CdTe at low temperatures around $10 \mu$m is very similar. Therefore, this provides some validation for the initial spectrometer design considerations of GREX-PLUS/HRS.

However, it is known that semiconductor materials like CdZnTe often exhibit variations in optical properties among batches, due to factors such as the amount of doped impurities. Thus, it is necessary to measure the changes in $n$ caused by differences among batches and the absence of birefringence. These properties would be indispensable to robustly develop an immersion grating for the mid-infrared.

\subsection*{Disclosures}
The authors declare that there are no financial interests, commercial affiliations, or other potential conflicts of interest that could have influenced the objectivity of this research or the writing of this paper.

\subsection*{Data}
The data presented in this article are publicly available in zenodo at https://zenodo.org/records/13905068.

\subsection* {Acknowledgments}
This work is supported by Grant-in-Aids for Scientific Research (Grant Number JP23H05441, Grant Number JP23H01222, and Grant Number JP23H01222) and also supported by JST SPRING (Grant Number JPMJSP2104).
We also thank the members of the Advanced Machining Technology Group at the JAXA Institute of Space and Astronautical Science for their kind help in manufacturing the metal parts required for the experiment. We would also like to express our gratitude for using penelope.ai to verify the structure of the paper.

\bibliography{report}   
\bibliographystyle{spiejour}   


\vspace{2ex}
\noindent\textbf{First Author} is a PhD student at the Graduate University for Advanced Studies (SOKENDAI) / Institute of Space and Astronautical Science (ISAS), JAXA. Her current research areas include infrared astronomy, optics, and protoplanetary disks.

\vspace{1ex}
\noindent Biographies and photographs of the other authors are not available.

\listoffigures
\listoftables

\end{spacing}
\end{document}